\documentclass[final, 10pt, twocolumn, conference]{IEEEtran}

\usepackage{color}
\usepackage{url}
\usepackage{blindtext, graphicx}
\usepackage{times}
\usepackage{graphicx}
\usepackage{amsmath, amssymb,latexsym}
\usepackage{url}
\usepackage{paralist}
\usepackage{epstopdf}
\usepackage{texdef2015}

\newtheorem{proposition}{Proposition} 
\newcommand{\Tcal}{\mathcal{T}}
\newcommand{\age}{\Delta}
\newcommand{\xseq}{\mathrm{x}}

\begin{document}
% \title{Status Age Analysis of Streaming Source Coding}
\title{Timeliness in Lossless Block Coding}
%{\large \textbf{Timeliness in Lossless Block Coding}}

\author{%
Jing Zhong and Roy D. Yates\\[0.5em]
{\small\begin{minipage}{\linewidth}\begin{center}
\begin{tabular}{ccc}
WINLAB, ECE Department\\
Rutgers, the State University of New Jersey\\
\{\emph{jzhong, ryates}\}@winlab.rutgers.edu
\end{tabular}
\end{center}\end{minipage}}
}

\maketitle
\thispagestyle{empty}

\begin{abstract}
\boldmath We examine lossless data compression from an average delay perspective.
An encoder receives input symbols one per unit time from an i.i.d. source and submits binary codewords to a FIFO buffer that transmits bits at a fixed rate to a receiver/decoder. Each input symbol at the encoder is viewed as a status update by the source and the system performance is characterized by the status update age, defined as the number of time units (symbols) the decoder output lags behind the encoder input.  
% We applied the status age analysis to real-time lossless source coding system in which source symbols are encoded sequentially and sent to the interested recipient through an error-free channel, and showed that the timeliness metric is strongly connected to the end-to-end delay of the coding system. 
An upper bound on the average status age is derived from the exponential bound on the probability of error in streaming source coding with delay. Apart from the influence of the error exponent that describes the convergence of the error, this upper bound also scales with the constant multiplier term in the error probability. 
However, the error exponent does not lead to an accurate description of the status age for small delay and small blocklength.
An age optimal block coding scheme is proposed based on an approximation of the average age by converting the streaming source coding system into a D/G/1 queue.
% Exploiting the quasi-linear property of the average age expression in block coding, we also proposed the age-optimal block coding scheme that minimizes the average age.
We compare this scheme to the error exponent optimal coding scheme which uses the method of types. We show that maximizing the error exponent is not equivalent to minimizing the average status age.

% \boldmath We examine lossless data compression from an average delay perspective. An encoder receives input symbols one per unit time from an i.i.d. source and submits binary codewords to a FIFO buffer that transmits bits at a fixed rate to a receiver/decoder. Each input symbol at the encoder is viewed as a status update by the source and the system performance is characterized by the status update age, defined as the number of time units (symbols) the decoder output lags behind the encoder input.  

% (incomplete)

% Encoding large blocks can result in high status age because the resulting arrival process at the FIFO buffer is bursty with long codewords that arrive infrequently. On the other hand, small blocklength encoding yields inefficient compression, which increases the offered load on the FIFO buffer and causes queueing delays. We characterize this status age tradeoff in terms of source entropy, encoder block length and FIFO buffer rate.
\end{abstract}


% \begin{IEEEkeywords}
% source coding, queue-theoretic system
% \end{IEEEkeywords}

\section{Introduction}

In this era of ubiquitous connectivity and computing with mobile devices, \emph{real-time status} updates ensure that a monitor (receiver) stays current about the status of interest of the source. This requires the status updates to be as \emph{timely} as possible. 
In \cite{Kaul2012,Yates2012}, a new delay metric, the \emph{average status age}, was introduced to measure the end-to-end timeliness of a status updating system. 
In the context of updates delivered through queues, general expressions of the average status age have been derived for single and multiple sources. 
In \cite{Kam2013,Costa2014,Huang2015}, status age analysis has been also applied to other communication systems, including random networks that deliver packets out of order and multi-class queueing systems.

% It is shown that the goal of timely updating is not equivalent to maximizing the utilization of the communication system that delivers the updates. 
% Instead, an optimal updating strategy balances between the load of the communication system and the frequency of status update messages.
% The status-age analysis of a more general cloud based system, where the status update messages are delivered out of order via a network cloud, is discussed in \cite{Kam2013}.

Many real-time data compression and communication systems with low latency requirements can be modeled as status updating systems in which the applications require the source to be reconstructed at the decoder in a timely manner.
These range from real-time video surveillance to remote telesurgery \cite{Butner2003}. 
The analysis of these timely compression and communication problems can be simplified to a real-time source coding problem over a data network.
% Source video frames are generated and revealed to the encoder in real-time, get encoded into bit sequences and delivered to the decoder for reconstruction through the communication system.
% In video streaming applications, source video frames are generated and revealed to the encoder in real-time. 
% These applications require the source to be reconstructed at the decoder as timely as possible. 
% For instance, if the video frames received by the doctor are outdated by a second, it is too risky for the doctor to take any actions since the circumstance of the patient has changed during the network latency.
The timely decoding of messages at receiver must balance data compression delays against network congestion deriving from insufficient compression.

Streaming source coding with a delay constraint was first discussed in \cite{Chang2006} and \cite{Chang2007}, in which the decoding error probability is bounded exponentially as a function of the delay constraint. 
In \cite{Draper2014}, the error exponent analysis is generalized to distributed streaming sources, and an achievable error exponent is obtained  by fixed rate coding schemes using random binning. 
Compared to this previous work, we are interested in the following question: how timely can the streaming source coding system be if we are allowed to choose any lossless fixed-to-variable block coding scheme? 
We approach this question by first connecting the status update age to the error exponent in lossless streaming source coding. 
Although the error analysis provides an upper bound on the timeliness measure, we will see that maximizing the exponent does not optimize the timeliness of decoded source messages.

In this work, we start in Section 2 with the system model of lossless streaming source coding problem, and derive an expression of average status age in lossless block coding schemes.
In Section 3, we show the upper bound of the average status age as a function of the error exponent.
% For fixed-to-variable length coding schemes, we find the upper and lower bounds and a good approximation for the average status age of the system, given a fixed block length and the probabilities of different codeword lengths. 
We use an example of block coding scheme to demonstrate how the constant term in the error probability leads to the difference between the bound and the actual average status age.
We then propose a method to find the age-minimizing optimal block coding scheme for average age in Section 4. We show in Section 5 that this is generally differs from the error-exponent maximizing coding scheme that uses the method of types.
We conclude with a summary of our work and possible future extensions in Section~6. 

% \section{Previous Work on Streaming Source Coding}

\section{System Model}

\begin{figure*}[t]
\centering
\includegraphics[width=.9\textwidth]{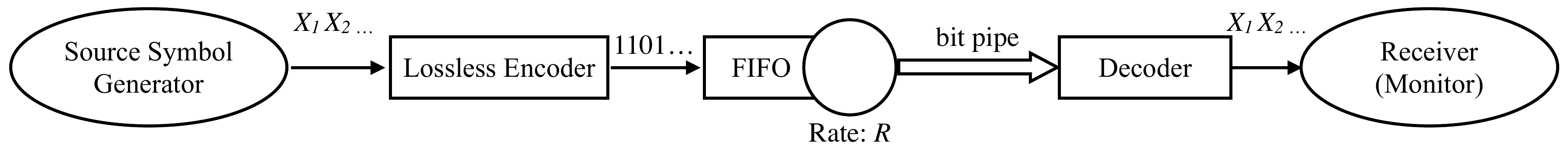}
\caption{System diagram of streaming source coding controlled by the FIFO buffer.}
\label{fig:source_sys}
\end{figure*}

The streaming source coding system introduced in \cite{Chang2006} and \cite{Chang2007} is illustrated in Figure \ref{fig:source_sys}. 
We assume that the channel between the encoder and decoder is a constant rate bit pipe with zero propagation delay. 
Starting at time $t=1$, discrete memoryless source symbols with finite alphabet $\mathcal{X}$ arrive at each time unit sequentially, so the source symbol $X_i$ arrives at time $i$.
In this work, we focus on fixed-to-variable length block coding schemes. %assuming that the encoder has full knowledge of the source symbol distribution. 
The encoder groups every $B$ message symbols into a single block and maps entire blocks into variable-length bit strings. 
The $k^{\mathrm{th}}$ symbol block is $Y_k$ such that $Y_{k+1} =  X_{kB+1} X_{kB+2} \cdots X_{(k+1)B} $.
The encoded sequence is then fed into a first-in-first-out (FIFO) buffer, which outputs one binary bit to the decoder through the channel every $1/R$ seconds. If the buffer is empty, it outputs a gibberish bit $e$ independent of any codewords. In fixed-to-variable length coding, the decoder is able to determine whether the next received bit is a gibberish bit or not, since the generation time of next symbol block is known to the decoder \cite{Chang2006}. 

When the decoder receives the prefix-free codeword, it reconstructs the corresponding message block immediately. The delivery time of the block $Y_k$ is denoted by $D_k$. Note that all the message symbols contained in a single message block are decoded at the same time and thus have the same delivery times. 

In the source coding problem, the status age is defined as the age of the most recently decoded symbol from when that symbol was generated.
That is, if the most recently decoded symbol at time $t$ is $X_i$, which was produced by the source at time $i$, the instantaneous status age is $\age(t) = t - i$. 
We observe that $\age(t)$ is a random process that varies in time with the receiver's reconstruction of the source.
The time-average status age of the coding system observed by the receiver is given by
\begin{align}
\age = \lim_{\Tcal\rightarrow\infty} \frac{1}{\Tcal} \int_0^\Tcal \age(t) \, \mathrm{d}t \;. 
\end{align}

\begin{figure}[t]
\centering
\includegraphics[width=0.5\textwidth]{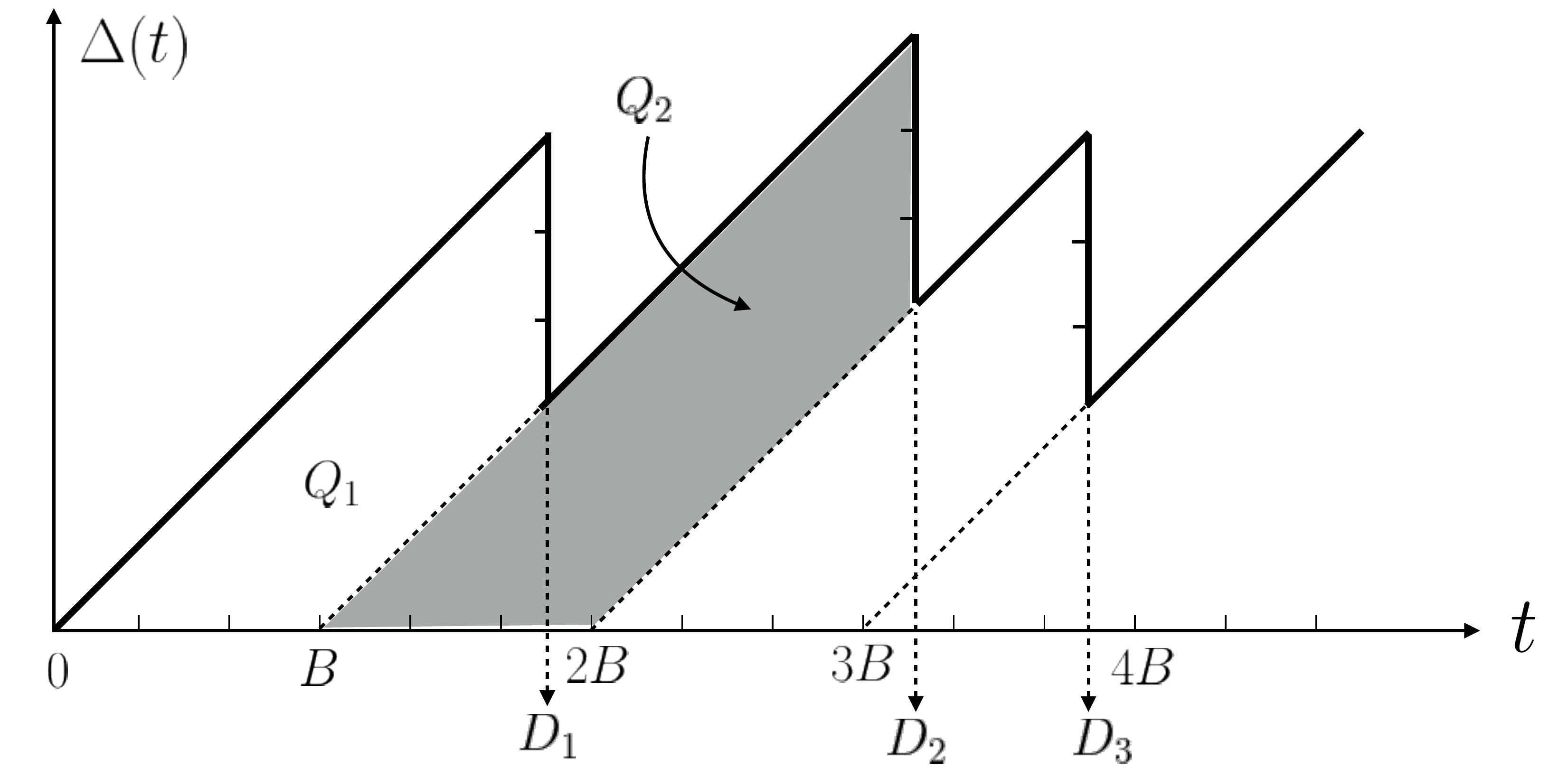}
\caption{Example of variation in status age at a receiver of streaming fixed-to-variable length coding with blocklength $B=3$.}
\label{fig:sawteeth}
\end{figure}

In streaming block coding, the status age at time $t$ is the same as the age of the last symbol in the most recently decoded block, since all symbols in the same block are decoded simultaneously. 
% and we define $D_k$ as the delivery time of that same block at the receiver. 
Fig.~\ref{fig:sawteeth} shows a sample realization of status age, as a function of time, at the receiver. 
We observe that $kB$ is the arrival time of symbol block $Y_k$ at the input of encoder. 
The age is a sawtooth function that increases linearly in time in the absence of any symbol blocks and is reset to $D_k-kB$ at time $D_k$ when symbol block $Y_k$ is decoded at the receiver. 

Using the same approach as \cite{Kaul2012}, the integration of the sawtooth area is equivalent to the sum of disjoint polygon areas $Q_k$ shown in Fig. \ref{fig:sawteeth}.  
The average status age for block coding can be expressed as
\begin{align}
\age & = \lim_{N\rightarrow\infty} \frac{1}{BN} \sum^N_{k=1} Q_k, \label{sum_saw} \\
% & = \lim_{N\rightarrow\infty} \frac{1}{BN} \sum^N_{k=1} \frac{[D_k-(k-1)B]^2}{2} - \frac{[D_k-kB]^2}{2} \\
& = \lim_{N\rightarrow\infty} \frac{1}{N} \sum_{k=1}^{N}(D_k-kB) + \frac{B}{2}. \label{age_blockcode}
\end{align}
From a queueing perspective, we can view a block $Y_k$ as a customer arriving at time $kB$ and departing at time $D_k$. The system time of this customer is $D_k-kB$.
Since $D_k\geq kB$, the average status age is always lower bounded by $B/2$. 
%The status age process is the sum of shifted non-overlapping triangles if the FIFO service rate is infinitely large and every symbol block is served and transmitted immediately in the queue.
% Given the average age expression (\ref{age_blockcode}) for fixed-to-variable length coding, the status age analysis of block coding can be transformed into a queueing delay problem. 
% If we group the encoder-buffer-decoder component as a black box, this black box is equivalent to a queue whose customers are the arriving source symbol blocks.
The interarrival times of customers are deterministic since the interarrival time of symbol blocks is exactly the block length $B$.
As a result, the term $\Eop[D_k-kB]$ in (\ref{age_blockcode}) is the expected system time that block $Y_k$ spends in the queue. Let $\Eop[T] \triangleq \Eop[D_k-kB]$, for an arbitrary customer $k$ when the queue has reached steady-state. 
Thus,
\begin{align}
\age = \Eop[T] + \frac{B}{2}. \label{age_blockcode2}
\end{align}
Intuitively, the timeliness metric $\Delta$ is translated into a end-to-end average delay measure in block coding schemes.
In (\ref{age_blockcode2}) the expected system time in a queue is given by the sum
\begin{align}
\Eop[T] = \Eop[S] + \Eop[W], \label{system_sum}
\end{align}
where $\Eop[S]$ and $\Eop[W]$ are the expected service time and the expected waiting time. 
In streaming block coding, each encoded bit takes $1/R$ time unit to be transmitted by the FIFO buffer, thus the service time of the symbol block $Y_k$ with corresponding binary code length $L_k$ is $S_k = L_k/R$, and $\Eop[S] = \Eop[L]/R$. 
Applying the upper bound for G/G/1 queue in \cite{Marshall1968}, the expected waiting is upper bounded by
\begin{align}
\Eop[W] \leq \frac{\Eop[L^2]-\Eop^2[L]}{2R(BR-\Eop[L])}.
\end{align}
Thus the average status age is upper bounded by 
\begin{align}
\age \leq \frac{\Eop[L^2]-\Eop^2[L]}{2R(BR-\Eop[L])}+\frac{\Eop[L]}{R} + \frac{B}{2}. \label{up}
\end{align}

\section{Connecting Status Age to Error Exponent}

In \cite{Chang2007thesis}, the traditional block coding error exponent is generalized to a streaming source coding problem, and the delay-constrained error exponent is introduced to describe the convergence rate of the symbol-wise error probability as a function of the decoding delay. 
At time $n$, the decoder estimates the $k^\mathrm{th}$ source symbol as $\hat{\xseq}_k(n)$, for all $k<n$. 
A delay constrained error exponent $E_s(R)$ is said to be achievable if and only if for all $\epsilon > 0$ and decoding delay $\delta > 0$, there exists a constant $K < \infty$ for a fixed-delay $\delta$ encoder-decoder pair such that
\begin{align}
\Pr [\hat{\xseq}_{n-\delta}(n) \neq \xseq_{n-\delta}] \leq K 2^{(-\delta E_S(R)-\epsilon)}, \qquad \textrm{for all }n>\delta. \label{error_exp}
\end{align}

In a lossless block coding system, a symbol block is successfully decoded only after the entire encoded bit sequence corresponding to that block departs from the FIFO buffer. An error occurs at time $n$ if some queueing delays cause some encoded bits of the symbol $n-\delta$ to be still in the buffer. 
The exponential convergence rate of error in delay $\delta$ comes from the randomness of the lengths of encoded bit sequences. 
% However, this is not true for asymptotically lossless coding schemes like random binning.

\begin{proposition}
A block coding scheme with achievable error exponent $E_S(R)$ has average status age $\age$ satisfying
\begin{align}
\Delta \leq \frac{K \, 2^{2E_S(R)}}{(2^{E_S(R)}-1)^2} \triangleq f(K,E_S(R)). \label{age_ee}
\end{align}
\end{proposition}
Proof of this proposition is shown in Appendix B. 
Despite the fact that the constant $K$ may vary for different $E_S(R)$, we observe that as $E_S(R)~\rightarrow~\infty$, $f(K,E_S(R))~\rightarrow~K$. 
This tells us both the error exponent and the constant term $K$ influences the status age in a streaming coding system. 
Although the exponent $E_S(R)$ provides a tight upper bound on the error probability when the delay constraint $\delta$ is large enough, it does not accurately describe the complicated error events when $\delta$ is small.
In the following discussion, we will use a prefix block code example to show the effect of $K$ and explain why an exponential bound is sometimes not good enough for the description of delay.

The following simple example was used in \cite[Sec.~2.2.2]{Chang2007thesis} to demonstrate a non-asymptotic error exponent result for a prefix-free block code.
Consider a source with alphabet size 3 and distribution
$P_X(A) = a, \; P_X(B) =  P_X(C) = (1-a)/2,$
where $a\in[0,1]$. Assume the encoder has no information about the source distribution, and chooses a block encoding strategy with blocklength $B=2$ as follows:
% \begin{align*}
% & AA \rightarrow 0, \\
% & AB \rightarrow 1000, \; AC \rightarrow 1001, \;
% \end{align*}
It maps the block $AA\rightarrow 0$ and all other blocks to 4-bit sequences led by a single 1, e.g., $AB \rightarrow 1000$. 
This coding scheme is not adapted to the source distribution, but it was shown to provide a convenient way to obtain a closed form expression of the error probability. 
It is also assumed that the channel rate is $R=3/2$, meaning that the FIFO buffer outputs 3 bits every 2 symbol periods. 
The average age $\age$ is finite if and only if the average length of codeword is less than the channel rate $BR$, i.e. $a^2+4(1-a^2)<3$. That is, $a>1/\sqrt{3}$ or $q>1/3$ if we define $q=a^2$.

Compared to \cite{Chang2007thesis} that seeks an exponential upper bound for the error probability, the calculation of (\ref{age_ee}) requires an exact expression including the constant $K$.
The error probability of this coding system is upper bounded by
\begin{align}
P_e \leq \left\{\begin{matrix}
1 \, , & 0\leq\delta<1 \\ 
1-q+\eta^{-3/2}[q(1-\eta)] 2^{\delta (\frac{3}{2}\log_2 \eta)}, & 1\leq\delta<3 \\ 
\eta^{-9/2}[1-q(1-\eta^3)] 2^{\delta (\frac{3}{2}\log_2 \eta)}, & \delta\geq 3
\end{matrix}\right. \label{pe_example}
\end{align}
where $\eta = \frac{-1+\sqrt{1+\frac{4(1-q)}{q}}}{2}$. The detailed procedure to obtain $P_e$ is described in Appendix B.

We observe in (\ref{pe_example}) that the achievable error exponent that satisfies  (\ref{error_exp}) is $-\frac{3}{2}\log_2 \eta$. 
However, the exponential bound is loose for small delay, specifically $0\leq\delta<3$. 
Intuitively, the decoder is still waiting for the binary codeword when $0\leq\delta<1$, so the probability is always upper bounded by 1 for any decoding strategy; when $1\leq\delta<3$, the decoder can successfully recover the message if the codeword is 0 and the buffer delivers this bit no later than $\delta$; when $\delta\geq 3$, we have the most general case that can be described by the exponential bound.
The error exponent analysis usually requires $\delta$ to be large enough, but the accumulation of average age also counts small values of $\delta$ that lead to large probability of error.
Following the steps of the proof of Proposition 1 in Appendix A, we express the upper bound on the average status age as
\begin{align}
\age & \leq 8 - 3q\left(1-\eta^{\frac{3}{2}}+\eta^{\frac{5}{2}}+\frac{2\eta}{3}\right) \nn
& \qquad + \frac{3K2^{-2E_S(R)}}{2^{E_S(R)}-1} + \frac{K2^{-E_S(R)}}{(2^{E_S(R)}-1)^2} , \label{prefix_up}
\end{align}
where $K=\eta^{-9/2}[1-q(1-\eta^3)]$ is the constant term from the exponential bound when $\delta\geq 3$ in (\ref{pe_example}).

\begin{figure}[t]
\centering
\includegraphics[width=0.5\textwidth]{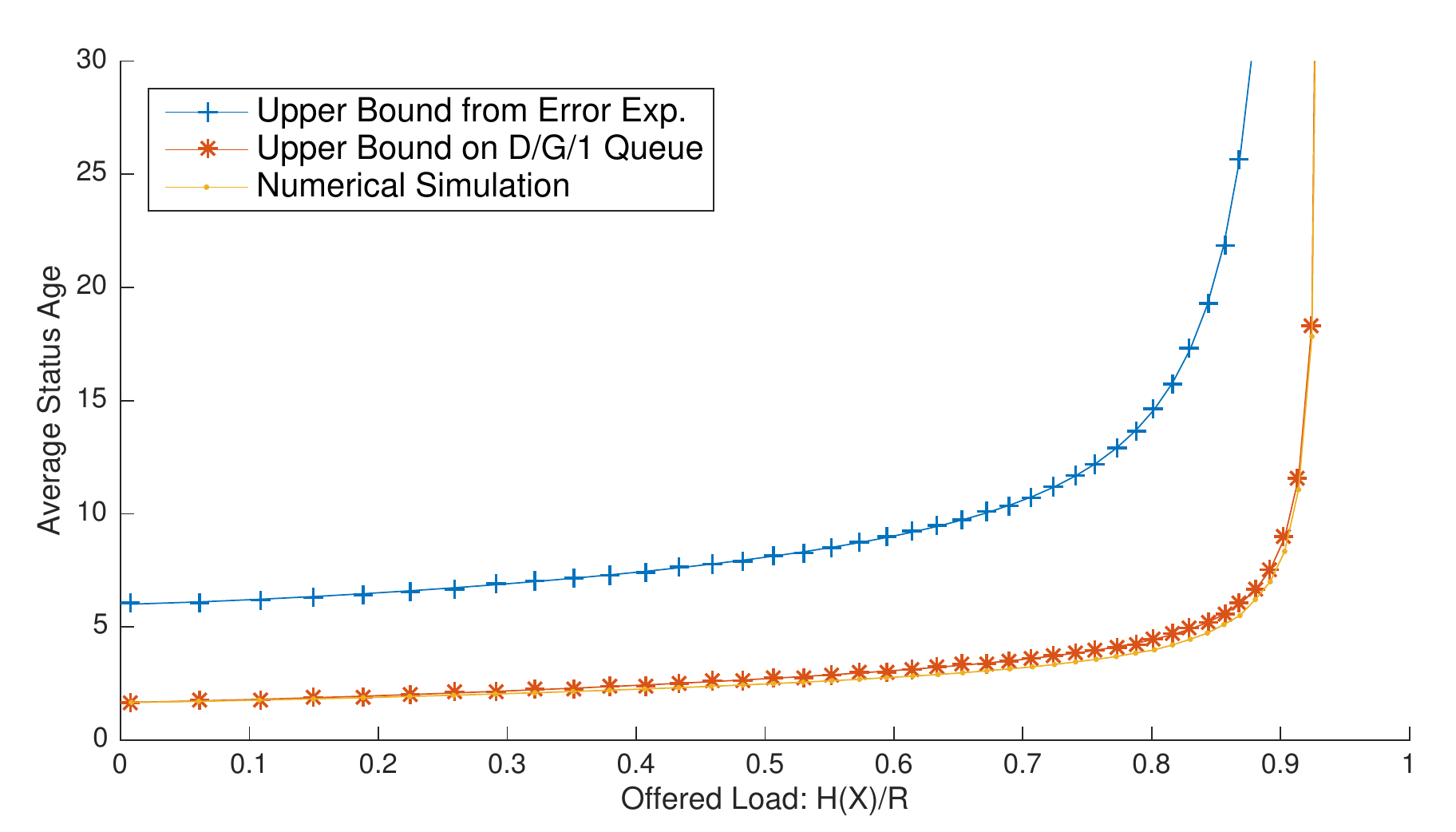}
\caption{Sample average age for prefix block code with blocklength $B=2$.}
\label{fig:prefix_ex}
\end{figure}

Figure \ref{fig:prefix_ex} depicts a comparison among the numerical simulation of average age, the upper bound obtained from D/G/1 queue in (\ref{up}) and the upper bound obtained from error exponent in (\ref{prefix_up}). 
In this plot, the channel rate $R$ is fixed at $3/2$, and the symbol probability $a$ is varied within $(1/\sqrt{3},1]$ such that the entropy $H(X)$ is also varied. 
The sharp transition effect occurs as the offered load of the system approaches 1 for all curves. 
This is because the average codeword length approaches $R$, and the number of bits queued in the FIFO buffers becomes unbounded. 
This effect occurs earlier than $H(X)/R=1$ since the coding scheme is not adapted to the source distribution.
We observe that the bound obtained from D/G/1 waiting time (\ref{up}) is tight to the true simulation, while the delay-constrained error exponent provides only a loose characterization of status age in a block coding system. 
As $a\rightarrow 1$, i.e. $H(X)\rightarrow 0$, the upper bound obtained from error exponent (\ref{prefix_up}) is dominated by the constant terms come from the sum of high error probabilities in the small $\delta$ region.

\section{Age Optimal Block Code}

\begin{figure}[t]
\centering
\includegraphics[width=0.45\textwidth]{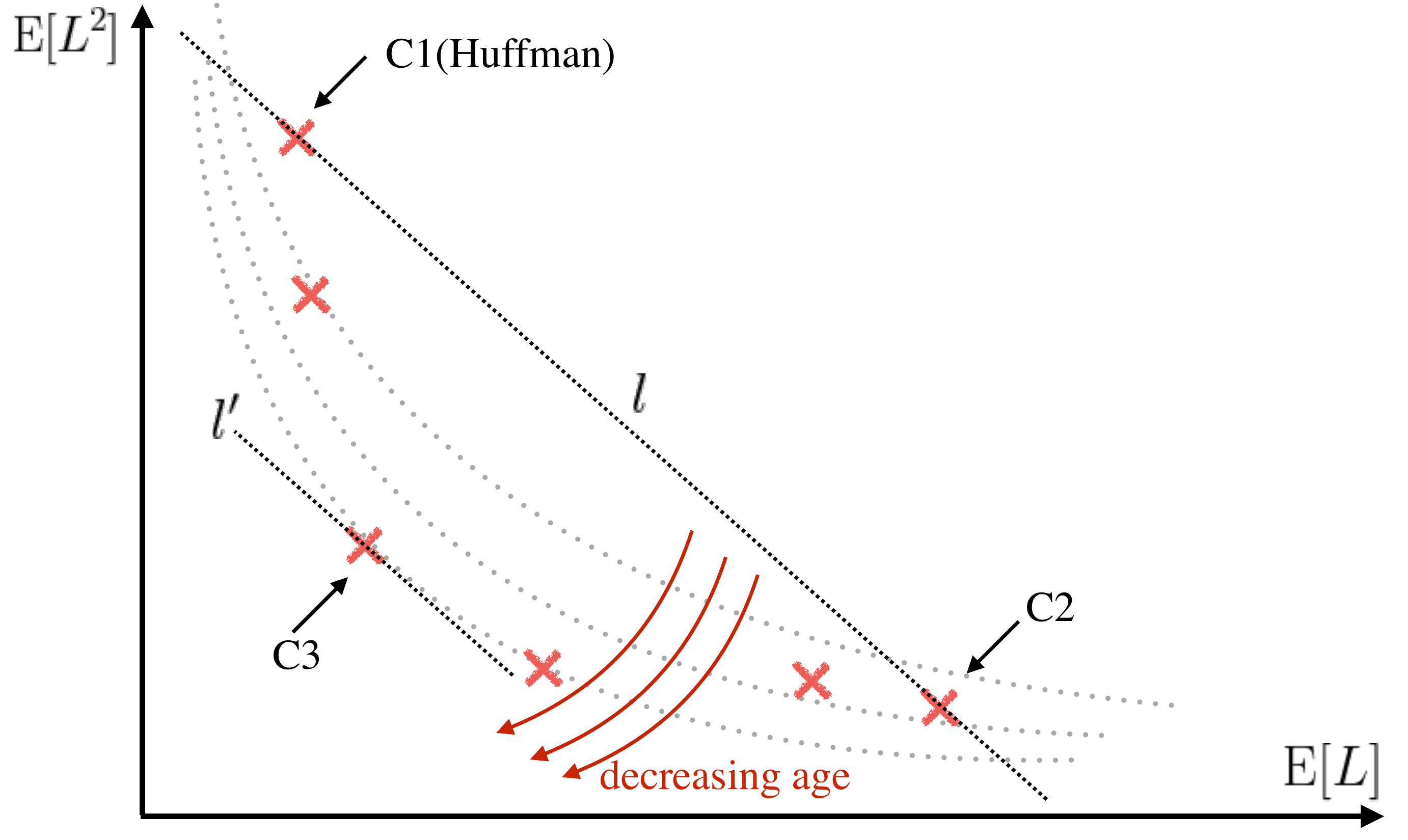}
\caption{The illustration of convex hull algorithm and the representation of code trees in the coordinate.}
\label{fig:hull}
\end{figure}

A block coding scheme is \emph{age-optimal} if it minimizes the average status age for a given $B$ and $R$. 
Since the bound in (\ref{up}) is simple and reasonablly tight, we use it as an approximation of the average status age and treat it as a penalty function with respect to variable $L$. 
Figure \ref{fig:hull} depicts a graphical representation of all the possible codebooks in the two dimensional space constructed by $\Eop[L]$ and $\Eop[L^2]$. 
It is proved in \cite{Larmore1989} that the set of all possible codebooks forms a convex hull for block coding schemes, and a linear approximation algorithm is introduced to iteratively search all code trees lying on the lower left boundary of the convex hull.  
The non-linear penalty function in (\ref{up}) is approximated by a linear function 
\begin{equation*}
f(L)=\alpha \Eop[L] + \beta \Eop[L^2], 
\end{equation*}
and it is assumed there is an efficient algorithm \emph{Find\_best}$(\alpha,\beta)$ which returns the codebook that minimizes the penalty function $f(L)$ given any $\alpha,\beta\in[0,1]$.
The algorithm starts from two extreme cases: $(\alpha,\beta)=(1,0)$ and $(\alpha,\beta)=(0,1)$. Note $(\alpha,\beta)=(1,0)$ is the penalty function that returns a code $C_1$ that is a Huffman code. Furthermore, $(\alpha,\beta)=(0,1)$ returns the minimum second moment code $C_2$. 
Given any two codebooks $C_1$ and $C_2$, new values are assigned to $\alpha$ and $\beta$ as follows:
\begin{align}
\alpha' & = \Eop[L^2](C_1) - \Eop[L^2](C_2) \\
\beta' & = \Eop[L](C_2) - \Eop[L](C_1) .
\end{align}
Afterwards, \emph{Find\_best}($\alpha',\beta'$) is called to search for a possible codebook between the points of $C_1$ and $C_2$.
Intuitively, this operation works as follows: we first draw a line segment $l$ that connects the points $C_1$ and $C_2$, and then \emph{Find\_best} identifies the lowest line $l'$ parallel to $l$ that meets the boundary of the convex hull of all codebooks. If $l'$ lies below $l$, then $l'$ contains at least one new codebook $C_3$. 
This step repeats iteratively by renewing the value of $\alpha$ and $\beta$ at each step for the $C1-C3$ and $C3-C2$ line segments, until we find all the feasible code trees.
A detailed description of the algorithm can be found in \cite{Larmore1989}.

Then the problem becomes how to efficiently return the best codebook given a linear penalty function $f(L)$. In \cite{Baer2006}, it is shown that the problem of source coding for linear penalties can be reduced to a coin collector's problem, and a recursive Package-Merge algorithm is introduced to solve the problem in linear time. 

Figure \ref{fig:optage_compare}(a) depicts a numerical example of the average age using age optimal code with different blocklength $B$. 
We use the three symbols $A, B, C$ with $P(A)=0.6, P(B)=~0.3$ and $P(C)=0.1$. And the channel rate $R$ is varied above $H(X)$ to vary the offered load. 
When $R$ is large compared to $H(X)$, the average age grows almost linearly with the blocklength $B$.   
Encoding with large blocklength is a losing proposition, since what we gain by reducing the output rate of the encoder is forfeited because the delay of the system is dominated by long interarrival times of large blocks.
% The arrival process at the FIFO buffer is bursty with long codewords that arrive infrequently.
Hence, the optimal strategy in the high FIFO rate region is to choose the smallest possible blocklength $B$. 
In contrast, as $R$ decreases, the sharp transition effect occurs earlier for smaller $B$ since the corresponding average code length is larger.
Since the redundancy of block coding decays with the blocklength $B$, the threshold of the transition approaches $\frac{H(X)}{R}=1$ as $B$ increases. 
We say that $B$ is a \emph{valid} blocklength for rate $R$ if and only if $R$ is larger than the code rate using blocklength $B$. 
In this region of transition, it is complicated to obtain the optimal blocklength analytically.

\section{Comparison to Optimal Error Exponent Code}

It is shown in \cite{Chang2007thesis} that the optimal error exponent $E_S(R)$ can be achieved by a prefix-free block coding scheme that uses the method of types.
For a message block of length $B$, the encoder first describes the type of the message block $\tau$ using $O(|\mathcal{X}|\log_2 B)$, then represent the index of the realization within this type by $B H(\tau)$ bits, where $H(\tau)$ is the entropy of the type.

\begin{figure}[t]
\begin{center}
\begin{tabular}{cc}
\includegraphics[width=0.45\textwidth]{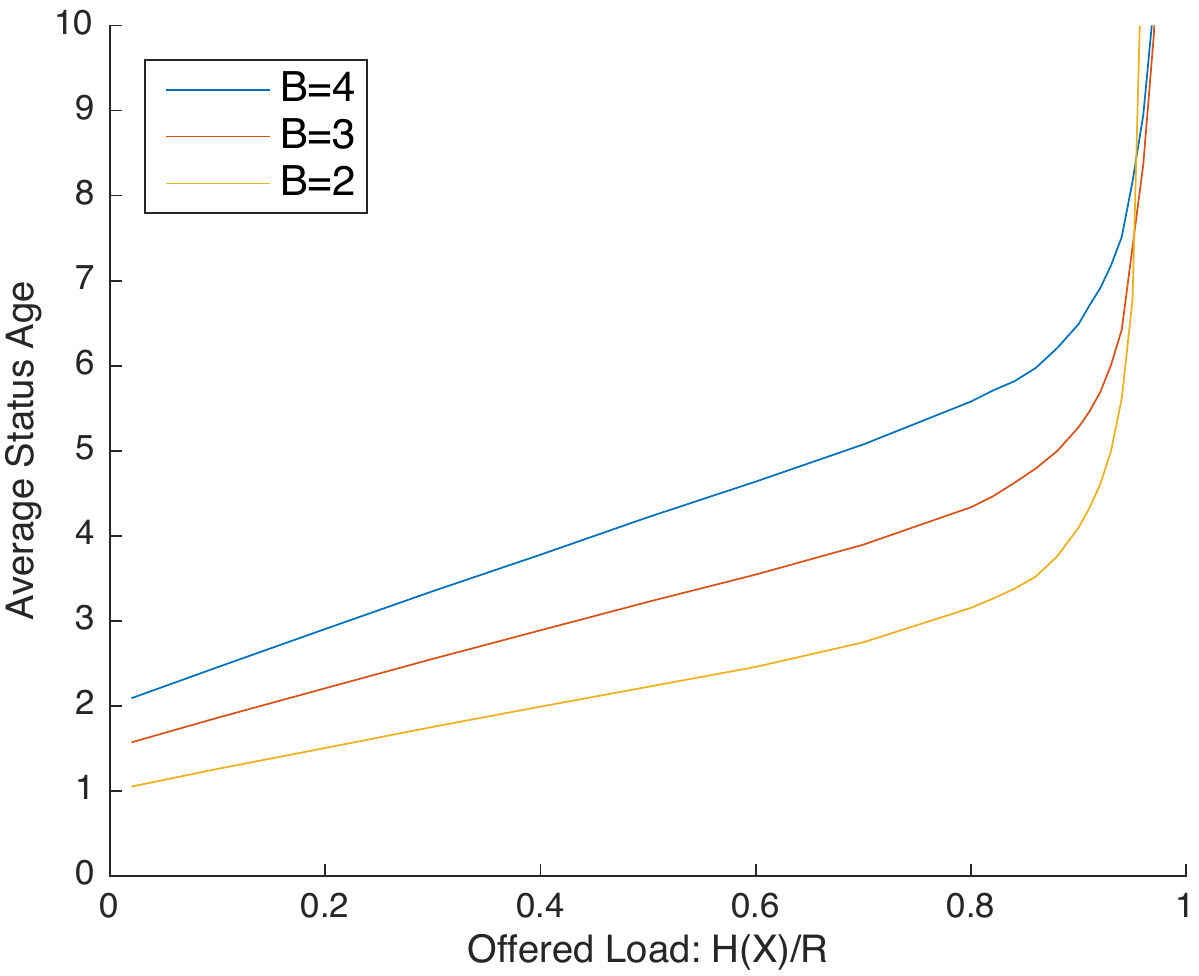}\\
{(a)} \\
\includegraphics[width=0.45\textwidth]{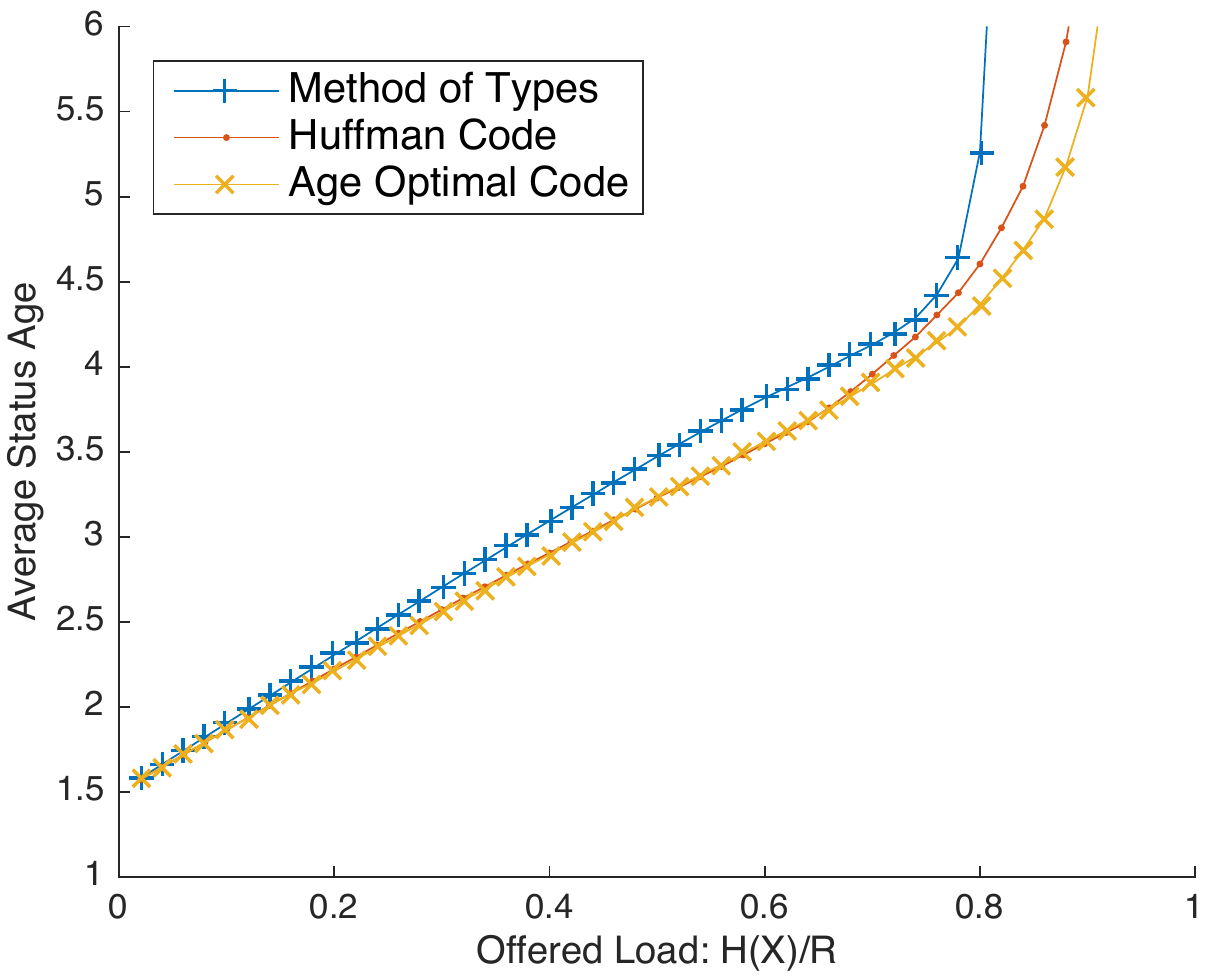}\\
{(b)}
\end{tabular}
\end{center}
\caption{Two numerical examples: (a) age optimal code with different blocklength $B$. (b) comparison between age optimal block code and the optimal error exponent code using method of types with blocklength $B=3$.}
\label{fig:optage_compare}
\end{figure}
Figure \ref{fig:optage_compare}(b) compares our age-optimal code to the coding scheme using method of types. We use the same source distribution as in Section 4 with blocklength $B=3$.
In this example, we observe that the age optimal code outperforms Huffman code when the offered load is high, implying that Huffman code is nearly optimal in the low load region.
Although the type coding scheme achieves the largest error exponent, it gives slightly larger average status age compared to the other two block coding schemes. 
This is because the type coding is asymptotically optimal in error, but the minimization of average age requires us to choose small blocklength since the term $B/2$ dominates in (\ref{age_blockcode}) when the channel rate $R$ is much larger than the source entropy $H(X)$.
% We also claim that any block coding schemes have the same codeword length distribution as the age-optimal block code if it provides the minimum average age for a given blocklength $B$. 

\section{Conclusion and Future Work}
We applied the status age analysis to a real-time lossless source coding system in which source symbols are encoded sequentially and sent to an interested recipient through an error-free channel, and showed that the timeliness metric is strongly connected to the end-to-end delay of the coding system. 
We connected the average age to the source coding error exponent with delay and discussed why the error exponent does not describe the delay in a non-asymptotic setup.
Exploiting the quasi-linear property of the average age expression in block coding, we also proposed the age optimal block coding scheme that minimizes the average age.
By comparing this scheme to the optimal coding scheme for error exponent which uses the method of types, we further showed that maximizing the error exponent is not equivalent to minimizing the average status age.

While we have focused here on toy examples, this work is a starting point for the application of status age analysis to real-time data compression. 
We examined how timely the streaming source coding system can be using lossless block coding schemes. 
We presented the connection between the age and error exponent numerically for small block length regime, although the asymptotic behavior remains unknown as the blocklength becomes large. 
A primary reason is that the exact expression of the constant $K$ of the error probability in (\ref{error_exp}) becomes too complicated when the blocklength gets large. 
Nevertheless, in practical settings of high speed networks, techniques for handling large blocklengths will be needed. 
Similarly, sources with memory and age-optimized universal coding schemes also merit attention. Finally, this work has shown that optimal real-time compression over a network depends strongly on the available network resources, even if the network is just a fixed-rate bit pipe. More realistic scenarios with shared network resources are also likely to be a rich source of unsolved problems. 

% is too complicated to obtain when the blocklength is large. 
% In practice, there should be a system constraint on the maximum blocklength since the average age grows linearly with the blocklength. From an engineering prospective, the maximum allowed blocklength will be depend on the maximum tolerable expected age.
% Furthermore, given the fact that the method of types achieves the optimal error exponent asymptotically in a universal way, we are also interested in the optimal universal code for the average age metric.

\section*{Appendix}

\subsection*{A. Proof of Proposition 1}

For any lossless coding schemes, the delivery time $D_k$ in (\ref{age_blockcode}) can also be defined as 
\begin{align}
D_k = \min \{ t | \hat{\xseq}^k(n) = \xseq^k, \textrm{for all } n\geq t \; \}.
\end{align}
Using the fact that $D_k-k$ only takes non-negative integer values, the differential time of the $k^{\mathrm{th}}$ symbol in the coding system can be written as 
\begin{align}
\E{D_k-k}  
& = \sum_{t=0}^{\infty} \Pr[D_k-k>t] \nn
& = \sum_{t=0}^{\infty} \Pr\{ \exists \hat{\xseq}^k(n) \neq \xseq^k, \textrm{for some } n\geq t+k \} \nn
& = \sum_{t=0}^{\infty} \prob{ \bigcup_{n\geq t+k} \xseq^k(n) \neq \xseq^k } .\label{union}
\end{align}
Following from the union bound of all the possible error events, we obtain
\begin{align}
\E{D_k-k}  
\leq \sum_{t=0}^{\infty} \sum_{n\geq t+k} \Pr[ \xseq^k(n) \neq \xseq^k] . \label{unionbound}
\end{align}
Note that for any lossless block codes in a point-to-point transmission system controlled by FIFO buffer, the error probability of the whole sequence is equivalent to the symbol-wise error probability since a source symbol is decoded only after all previous symbols were successfully decoded in advance. That is, $
 \Pr [\hat{\xseq}^{k}(n) \neq \xseq^{k}] = \Pr [\hat{\xseq}_{k}(n) \neq \xseq_{k}].$
Using the upper bound on the error exponent in (\ref{error_exp}), we have 
\begin{align}
\E{D_k-k}
\leq \sum_{t=0}^{\infty} K \sum_{\delta=t}^{\infty} 2^{-\delta E_S(R)}
% & = \sum_{t=0}^{\infty} \frac{K2^{(1-t)E_S(R)}}{2^{E_S(R)}-1} \\
= \frac{K 2^{2E_S(R)}}{(2^{E_S(R)}-1)^2} \; .
\end{align}
% In (\ref{event}) note that the event $\{D_k\leq t_k\} = \{\hat{\xseq}^k(n) = \xseq^k ,\textrm{for all } n\geq t_k \}$, and then $\{D_k> t_k\} = \{D_k\leq t_k\}^C = \{ \exists \xseq^k(n) \neq \xseq^k, \textrm{for some } n\geq t_k \}$. 
% From (\ref{union}) to (\ref{unionbound}), the union bound of all the possible error events is applied. In (\ref{sub_ee}) we use the upper bound on the error probability in (\ref{error_exp}).

\subsection*{B. Error Probability of Prefix Block Code Example.}

Let $B_k$ to be the number of bits in the buffer after the transition at time $2k$, then $B_k$ forms a Markov chain since the incoming codeword every two symbol time is either of length 1 or length 4. That is, $B_{k+1} = B_k -2$ with probability $q=a^2$, and $B_{k+1} = B_k +1$ with probability $1-q$. Note that the boundary condition is $B_k>0$. The stationary distribution of $B_k$ is obtained as 
\begin{align}
\mu_j = \Lambda \eta^j, \label{stationary}
\end{align} 
where $\Lambda$ is the normalizer such that $\sum_{j\geq 0} \mu_j = 1$ and $ \eta =-\frac{1}{2}+\frac{1}{2}\sqrt{1+\frac{4(1-q)}{q}}$. Thus we rewrite $\mu_j = \Lambda\eta^j = \eta^j-\eta^{j+1}$. The stationary distribution exists iff $\eta<1$, requiring $q\in(\frac{1}{3},1]$.
Denote $L_{k+1}$ as the next incoming codeword length after the buffer state $B_k$, and $\beta_\delta=\lfloor 3(\delta-1)/2\rfloor$. Following the stationary distribution in (\ref{stationary}), we can bound the symbol-wise error probability using the outage probability of the buffer and obtain
\begin{align}
P_{e} \; & \leq \;  \Pr[L_{k+1}=1] \prob{B_k>\beta_\delta-1} \nn
& \qquad \qquad + \Pr[L_{k+1}=4] \prob{B_k>\beta_\delta-4} \nonumber \\
& = \;  \left\{\begin{matrix}
1 \, , & 0\leq\delta<1 \\ 
1-q+\eta^{-3/2}[q(1-\eta)] 2^{\delta (\frac{3}{2}\log_2 \eta)}, & 1\leq\delta<3 \\ 
\eta^{-9/2}[1-q(1-\eta^3)] 2^{\delta (\frac{3}{2}\log_2 \eta)}, & \delta\geq 3.
\end{matrix}\right.
\end{align}

\bibliographystyle{IEEEtran}
\bibliography{ref}

\begin{thebibliography}{10}
\providecommand{\url}[1]{#1}
\csname url@samestyle\endcsname
\providecommand{\newblock}{\relax}
\providecommand{\bibinfo}[2]{#2}
\providecommand{\BIBentrySTDinterwordspacing}{\spaceskip=0pt\relax}
\providecommand{\BIBentryALTinterwordstretchfactor}{4}
\providecommand{\BIBentryALTinterwordspacing}{\spaceskip=\fontdimen2\font plus
\BIBentryALTinterwordstretchfactor\fontdimen3\font minus
  \fontdimen4\font\relax}
\providecommand{\BIBforeignlanguage}[2]{{%
\expandafter\ifx\csname l@#1\endcsname\relax
\typeout{** WARNING: IEEEtran.bst: No hyphenation pattern has been}%
\typeout{** loaded for the language `#1'. Using the pattern for}%
\typeout{** the default language instead.}%
\else
\language=\csname l@#1\endcsname
\fi
#2}}
\providecommand{\BIBdecl}{\relax}
\BIBdecl

\bibitem{Kaul2012}
S.~Kaul, R.~Yates, and M.~Gruteser, ``Real-time status: How often should one
  update?'' in \emph{Proc.\ INFOCOM}, Apr. 2012, pp. 2731--2735.

\bibitem{Yates2012}
R.~Yates and S.~Kaul, ``Real-time status updating: Multiple sources,'' in
  \emph{Proc. IEEE Int.\ Symp.\ Inform.\ Theory}, Jul. 2012, pp. 2666--2670.

\bibitem{Kam2013}
C.~Kam, S.~Kompella, and A.~Ephremides, ``{Age of information under random
  updates},'' in \emph{Proc. IEEE Int.\ Symp.\ Inform.\ Theory}, Jul. 2013, pp.
  66--70.

\bibitem{Costa2014}
M.~Costa, M.~Codreanu, and A.~Ephremides, ``{Age of information with packet
  management},'' in \emph{Proc. IEEE Int.\ Symp.\ Inform.\ Theory}, 2014, pp.
  1583--1587.

\bibitem{Huang2015}
L.~Huang and E.~Modiano, ``{Optimizing age-of-information in a multi-class
  queueing system},'' in \emph{Proc. IEEE Int.\ Symp.\ Inform.\ Theory}, Jun.
  2015, pp. 1681--1685.

\bibitem{Butner2003}
S.~Butner and M.~Ghodoussi, ``Transforming a surgical robot for human
  telesurgery,'' \emph{IEEE Trans.\ Robotics and Automation}, vol.~19, no.~5,
  pp. 818--824, Oct. 2003.

\bibitem{Chang2006}
C.~Cheng and A.~Sahai, ``The error exponent with delay for lossless source
  coding,'' in \emph{IEEE Inf.\ Theory Workshop}, Mar. 2006, pp. 252--256.

\bibitem{Chang2007}
C.~Chang and A.~Sahai, ``Delay-constrained source coding for a peak distortion
  measure,'' in \emph{Proc. IEEE Int.\ Symp.\ Inform.\ Theory}, 2007, pp.
  576--580.

\bibitem{Draper2014}
S.~C. Draper, C.~Chang, and A.~Sahai, ``{Lossless coding for distributed
  streaming sources},'' \emph{IEEE Trans.\ Inf.\ Theory}, vol.~60, no.~3, pp.
  1447--1474, 2014.

\bibitem{Marshall1968}
K.~Marshall and R.~V. Evans, ``Some inequalities in queuing,'' \emph{Operations
  Research}, pp. 651--668, 1968.

\bibitem{Chang2007thesis}
C.~Chang, ``{Streaming source coding with delay},'' Ph.D. dissertation, UC
  Berkeley, 2007.

\bibitem{Larmore1989}
L.~L. Larmore, ``{Minimum delay codes},'' \emph{SIAM Journal on Computing},
  1989.

\bibitem{Baer2006}
M.~B. Baer, ``Source coding for quasiarithmetic penalties,'' \emph{IEEE Trans.\
  Inf.\ Theory}, vol.~52, no.~10, pp. 4380--4393, 2006.

\end{thebibliography}

\end{document}